\documentclass[iop]{emulateapj-rtx4}
\usepackage{amsmath}
\usepackage{graphicx,graphics}
\usepackage{rotating}
\usepackage{natbib}

\newbox\grsign \setbox\grsign=\hbox{$>$} \newdimen\grdimen \grdimen=\ht\grsign
\newbox\simlessbox \newbox\simgreatbox
\setbox\simgreatbox=\hbox{\raise.5ex\hbox{$>$}\llap
     {\lower.5ex\hbox{$\sim$}}}\ht1=\grdimen\dp1=0pt
\setbox\simlessbox=\hbox{\raise.5ex\hbox{$<$}\llap
     {\lower.5ex\hbox{$\sim$}}}\ht2=\grdimen\dp2=0pt
\def\simgreat{\mathrel{\copy\simgreatbox}}

\def\bfnabla{{\mbox{\boldmath $\nabla$}}}

\renewcommand\bv{{\mbox{\boldmath $v$}}}

\newcommand\bP{{\mbox{\boldmath $P$}}}
\newcommand\bF{{\mbox{\boldmath $F$}}}
\newcommand\bfr{{\sf\boldmath f}}
\newcommand\bI{{{\sf\boldmath I}}}

\newcommand\Crat{\mathbb{C}}
\newcommand\Prat{\mathbb{P}}

\def\<{\,\langle\langle}
\def\>{\,\rangle\rangle}

\begin{document}

\title{The effects of irradiation on the cloud evolution in active galactic nuclei}

\shortauthors{D. Proga~at~al.}
\author{Daniel Proga\altaffilmark{1}, Yan-Fei Jiang\altaffilmark{2,3}, Shane W. Davis\altaffilmark{4},
  James M. Stone\altaffilmark{2} and Daniel Smith\altaffilmark{1}}
\altaffiltext{1}{Department of Astronomy,University of
  Las Vegas,Las Vegas, NV 89119, USA} 
\altaffiltext{2}{Department of Astrophysical Sciences, Princeton
University, Princeton, NJ 08544, USA}
\altaffiltext{3}{Harvard-Smithsonian Center for Astrophysics, 60 Garden St., Cambridge, MA 02138, USA}
\altaffiltext{4}{Canadian Institute for Theoretical Astrophysics. Toronto, ON M5S3H4, Canada} 

\begin{abstract}
We report on the first phase of our study of cloud irradiation.
We study irradiation  by means of numerical, two-dimensional time-dependent 
radiation-hydrodynamic simulations of a cloud irradiated by a strong radiation. 
We adopt a very simple treatment of the opacity, neglect photoionization
and gravity, and instead focus on assessing the role of the type 
and magnitude of the opacity on the cloud evolution.
Our main result is that
even relatively dense clouds that are radiatively heated 
(i.e., with significant absorption opacity) do not move as a whole instead 
they undergo a very rapid and major evolution in its shape, size and physical 
properties. In particular, the cloud and its remnants become optical thin 
within less than one sound crossing time and 
before they can travel over a significant distance  (a distance of a few 
radii of the initial cloud). We also found that
a cloud can be accelerated as a whole under quite extreme conditions, e.g.,
the opacity must be dominated by scattering. However, the acceleration
due to the radiation force is relatively small and unless the cloud
is optically thin the cloud quickly changes its size and shape. 
We discuss implications for the 
modelling and interpetation  broad line regions of active galactic nuclei.
\end{abstract}


\section{Introduction}
\label{sec:intro}

There are very many situations in astrophysics where one object or a
group of objects is exposed to relatively strong radiation 
produced by a nearby external source. Examples of such situations 
include planets and moons irradiated by their host star, 
a star irradiated by its companion in a binary system, 
gas clouds irradiated by a nearby stellar cluster or by
an active galactic nucleus (AGN), and finally 
an outer part of a flaring accretion disk irradiated by its inner part 
or by the accretor. 

Radiative heating caused by irradiation can change the irradiated object
in several ways, e.g., it could change its structure, shape, size
and the overall appearance. It  could also lead to a significant mass loss and 
even acceleration of the cloud away from the source of radiation 
(via the so-called rocket effect). 
To some degree, similar changes could be caused by the radiation force
(i.e., without radiative heating).

Effects of irradiation are most profound (e.g., cause ablation and destruction)
in cases where the radiation energy is relatively high and the mass of
the irradiated object is small so that self-gravity is negligible. 
Such cases are relevant in a variety of astrophysical
environments, e.g., in the interstellar medium (ISM) \citep[e.g.,][]
{OortSpitzer1955,Bertoldi1989,BertoldiMcKee1990,Bally1995}, 
in planetary nebulae \citep[e.g.,][]{Mellemaetal1998}, 
near the central region of AGN
\citep[e.g.,][and references therein]{Mathews1986}
and outside the AGN host galaxy, in the intergalactic
medium (IGM) \citep[e.g.,][]{DonahueShull1987}.

The key questions in this context include: 
How does the radiation that is reflected, reprocessed, or transmitted  
by a cloud compare to the original external radiation?
What aspects and to what degree can the observed structure and kinematics
be accounted for by irradiation?
What are the dominant processes responsible for
dispersing the gas that was initially collected in the cloud?
What are the dominant processes responsible for accelerating the cloud?
Can a cloud be significantly accelerated before it is dispersed?
How do the acceleration and dispersing rates compare with the cloud
formation rate?

Answering these questions is hard from both observational and theoretical 
points of view because the evolution of real clouds is very complex
and too slow to be measured directly by observations. In addition,
only in a few cases are clouds well spatially resolved (i.e.,
those in the nearby ISM, e.g., \cite[][]{Bally1995}). 
Proper modeling of the clouds is further complicated by the fact that many 
time-dependent and multi-demensional processes and effects are
involved, e.g., radiative transfer (RT), gas photoionization and heating, 
the subsequent development and propagation of ionization and thermal fronts
(IF and TF, respectively) and of shocks and discontinuities. 
Generally, the cloud irradiation problem requires the simultaneous solving of
radiation-hydrodynamic (R-HD) equations 
\citep[examples of such studies include][]{LeflochLazareff1994,
Mellemaetal1998,Gonzalezetal2007,Ragaetal2007}.
One important aspect of this problem is that  
irradiation is anisotropic and optically thick clouds will cast shadows.
Therefore, the RT methods for solving the R-HD equation
have to treat the shadows accurately
\citep[e.g.,][]{Davisetal2012,Jiangetal2012}.
Effects of magnetic fields and dust increase the level of complexity 
even more \citep[e.g.,][and references therein]{Krauseetal2012}.

\subsection{Simple Cases}
\label{sec:scases}

However, under some special and idealized conditions,
the solution to the problem of the  evolution of a spherical cold cloud 
that is
suddenly exposed to external radiation could be quite trivial.
For example, in optically thin cases with pure absorption opacity,
the radiation would propagate very fast (faster than sound) 
throughout a cloud. If the cloud was initially of constant density
and in pressure equilibrium with the uniform ambient medium, it
will be uniformily heated. Consequently, the cloud will
expand without changing its shape and without gaining net momentum (i.e.,
it will behave like an expanding balloon for weak irradiation or 
an exploding sphere for strong irradiation). 
We will refer to such cases as simple example~I.

Another class of trivial solutions exists for optically thin clouds with
pure scattering opacity. Then again the radiation would propagate
throughout the cloud very fast and afterward the cloud would
experience a uniform acceleration away from the radiation source.
The acceleration would be constant also with time (for 
small clouds far from the radiation source) and the cloud would gain
momentum without changing its size and shape and without mass loss
(i.e., in some respects, it will behave like a bullet).
We will refer to such cases as simple example~II.

Another simple yet non-trivial example, is when a cloud is very optically
thick due to absorption opacity and it is exposed to weak radiation. 
In such a case, the part of the cloud facing the radiation source
will be gently heated and the radiation will penetrate only a thin
layer of the cloud. The IF and TF would move very slowly across the cloud and
there would be no shock. In addition, the cloud would slowly lose mass from
its heated part and would be in a quasi-steady state. This simple case
and the early evolution of other special cases can be and have been
studied using analytic methods that have made various simplifying assumptions 
in order to estimate for example, the mass loss rate, the final velocity 
of the neutral cloud, and shape of the IF
\citep[e.g.,][]{OortSpitzer1955,Bertoldi1989}.

\subsection{Clouds in AGN}
\label{sec:agncl}

It is challenging to determine what type of clouds
are most relevant in a given environment. It is especially true
for the so-called Broad Line Regions (BLRs) 
in AGN because they are spatially unresolved. 
In most studies that aim at interpreting
or modeling the observed line emission and absorption produced 
in the BLRs, 
the cloud propriety -- such as the density, size, and shape --
have been assumed 
\citep[e.g.,][and references therein]{Mathews1974,BlumenthalMathews1975,BlumenthalMathews1979,Capriottietal1981,Mathews1982,AravLi1994,Krolik1999}.

The presence of broad emission lines (BELs) and broad absorption lines
(BALs) in AGN spectra shows that AGN continuum radiation affects 
the AGN's immediate environment.
BELs are one of the defining spectral features of AGN. They are observed
in optical and ultraviolet (UV) spectra and have line wings extending 
to velocities up to $10^4~{\rm km~s^{-1}}$.
It is well established that the primary physical mechanism
for production of BELs is photoionization by the compact
continuum source of AGN \citep[e.g.][and references
therein]{KwanKrolik1981,FerlandElitzur1984,
Ferlandetal1998,HamannFerland1999,Krolik1999}. Detailed
photoionization calculations presented in these and other studies yield
relatively tight constraints on some physical conditions
of the emitting gas (e.g., the gas temperature, 
$T_g\approx 10^4$~K, the number density, $n\approx10^9~{\rm cm^{-3}}$, 
the column density $N\simgreat 10^{22}~{\rm cm^{-2}}$, and the ionizing
flux, $F_{ion}$ is so high that
the ratio between the radiation to gas pressure $\Xi\equiv F_{ion}/c n k T\approx 1$).

The width of BELs indicates that the emitting gas is highly {\it supersonic}.
The shape and position of the BEL profiles has been traditionally explained
by lines emitted in a cloudy region without a preferred velocity direction
and with a nearly spherical distribution
\citep[e.g.,][and references therein]{UrryPadovani1995,Krolik1999}.
We note that another possibility is that the BEL are produced
at the base of a wind from an accretion disk 
\citep[e.g.,][]{Murrayetal1995,Bottorffetal1997}.

The key issues that any cloudy model for the BLRs faces are
stability and confinement of the clouds 
\citep[see][for reviews]{OsterbrockMathews1986,Krolik1999,Krauseetal2012}.
In short, the clouds in the BLRs
are  hydrodynamically unstable, the nature of their confinement 
is unclear, while the production of new clouds
appears to be inefficient and requires rather extreme conditions.
In addition, it has been argued that radiation would cause
significant shearing and destroy the clouds before they could
contribute to the line emission \citep[e.g.,][]{Krolik1988,MathewsDoane1990}.

As we mentioned earlier, modeling irradiated clouds is a very challenging
problem. Many previous studies of clouds in AGN used various simplifying 
assumptions and the robustness and applicability of their results remain quite
uncertain. Perhaps the  most robust results of the previous
work is that BELs in AGN are produced by  a 
photoionized and supersonic medium that is optically thick 
(i.e., $N > 10^{22}~{\rm cm^{-2}}$)
\citep[e.g.,][]{KwanKrolik1981,FerlandElitzur1984,SneddenGaskell2007,
Reesetal1989}.
If this medium is indeed made of optically thick clouds
then one must accurately treat RT, in particular, shadows.
This requires solving self-consistently time-dependent, 
multidimensional R-HD equations.

The direction of the flux is not known independently of the energy
density. Therefore,  methods based on the
diffusion approximation cannot represent shadows.  
This has been demonstrated through the irradiation
of very optically thick structures with a beamed radiation field 
\citep[e.g.,][]{HayesNorman2003,Gonzalezetal2007}.
However, there are other methods, like the variable Eddington tensor
(VET) method that can capture shadows accurately.

For example, \cite{Davisetal2012} have developed algorithms
to solve the RT equation using short characteristics to compute the VET.
The tests described in \cite{Davisetal2012} 
show the accuracy of the RT solver for two radiation beams
(see Figure 6 of that paper). 
This solver was implemented in the R-MHD code 
Athena \citep{Stoneetal2008,Jiangetal2012}.
\cite{Jiangetal2012} presented some results from a few test runs
of the dynamical evolution of a cloud ablated by
an intense radiation field. These tests show that the HD 
as well as MHD solvers implemented in Athena when coupled 
with the VET method capture shadows correctly. In particular,
by using an input radiation field at two angles,  \cite{Davisetal2012}
and  \cite{Jiangetal2012}   explored tests where
both umbra and penumbra are formed.  This makes the tests
more difficult, because ad-hoc closures that capture only one direction for
the flux will not represent both the umbra and penumbra correctly.

Here, using Athena we study the time evolution of clouds irradiated 
by a radiation field as strong as the field believed to irradiate the
gas in AGN (i.e., radiation pressure to be comparable to gas
pressure). We consider clouds with a range of properties: from
optically thin to thick and with the opacity due to scattering 
or absorption processes or both.
To explain BELs and BALs, clouds in AGN must
move with large velocities. Therefore, our primary focus
is on identifying physical conditions under which  a pre-existing 
cloud could be significantly accelerated before it is dispersed.

Athena allows us to explore the above mentioned wide
range of conditions. Nevertheless, our treatment of clouds is quite
simplified. For example, our simulations are in two-dimensions
(2-D) and we assume the clouds to be highly ionized. Therefore, 
we do not follow the development and evolution of the IF. 
Moreover, we  do not include dust grains and magnetic fields. 
These and other complications could be modeled with the code but are 
beyond the scope of this preliminary work.

The outline of this paper is as follows. We describe our calculations
in 
\S~\ref{sec:methods}. 
In \S~\ref{sec:results}, we present our results. 
We summarize our results and discuss them together with their
limitations in
\S~\ref{sec:summary}.

\section{Methods}
\label{sec:methods}
We solve the radiation hydrodynamic equations in the mixed frame with radiation source terms given 
by \cite{Lowrieetal1999}. We assume local thermal equilibrium (LTE) and that the Planck
and energy mean absorption opacities are the same. Detailed discussion of the equations we solve can be found in
\cite{Jiangetal2012}. The equations are
\begin{eqnarray}
\frac{\partial\rho}{\partial t}+\bfnabla\cdot(\rho \bv)&=&0, \nonumber \\
\frac{\partial( \rho\bv)}{\partial t}+\bfnabla\cdot({\rho \bv\bv+{\sf P}}) &=&-{\bf \tilde{S}_r}(\bP),\  \nonumber \\
\frac{\partial{E}}{\partial t}+\bfnabla\cdot\left[(E+P)\bv\right]&=&-c\tilde{S}_r(E),  \nonumber \\
\frac{\partial E_r}{\partial t}+\bfnabla\cdot \bF_r&=&c\tilde{S}_r(E), \nonumber \\
\frac{1}{c^2}\frac{\partial \bF_r}{\partial t}+\bfnabla\cdot{\sf P}_r&=&{\bf \tilde{S}_r}(\bP).
\label{dimequation}
\end{eqnarray}
Here, $\rho$ is density, ${\sf P}\equiv P\bI$ with $\bI$ the unit tensor 
and $P$ gas pressure, and $c$ is the speed of light.
The total gas energy density is
\begin{eqnarray}
E=E_g+\frac{1}{2}\rho v^2,
\end{eqnarray}
where $E_g$ is the internal gas energy density.   We adopt an equation of state
for an ideal gas with adiabatic index $\gamma$, thus
$E_g=P/(\gamma-1)$ and $T=P/R_{\text{ideal}}\rho$, where
$R_{\text{ideal}}$ is the ideal gas constant. 
The radiation pressure ${\sf P}_r$ is related to the radiation energy density $E_r$
by the closure relation
\begin{eqnarray}
{\sf P}_r=\bfr E_r.
\end{eqnarray}
where $\bfr$ is the VET, and $\bF_r$ is radiation flux. Finally, 
${\bf\tilde{S}_r}(\bP)$ and $\tilde{S}_r(E)$ are the radiation momentum 
and energy source terms, respectively.

Following \cite{Jiangetal2012}, we use a dimensionless set of equations and variables
in the remainder of this work.  We convert the above set of equations to the dimensionless 
form by choosing fiducial units for temperature, pressure, and velocity
as $T_0$, $P_0$, and $a_0=\sqrt{P_0/T_0}$, respectively. Then 
units for radiation energy density $E_r$ and flux $\bF_r$ are $a_rT_0^4$ 
and $ca_rT_0^4$. In other words, $a_r=1$ in our units. The dimensionless speed of light is $\Crat\equiv c/a_0$. 
The original dimensional equations can then be written in the following dimensionless 
form 
\begin{eqnarray}
\frac{\partial\rho}{\partial t}+\bfnabla\cdot(\rho \bv)&=&0, \nonumber \\
\frac{\partial( \rho\bv)}{\partial t}+\bfnabla\cdot({\rho \bv\bv+{\sf P}}) &=&-\mathbb{P}{\bf S_r}(\bP),\  \nonumber \\
\frac{\partial{E}}{\partial t}+\bfnabla\cdot\left[(E+P)\bv\right]&=&-\mathbb{PC}S_r(E),  \nonumber \\
\frac{\partial E_r}{\partial t}+\mathbb{C}\bfnabla\cdot \bF_r&=&\mathbb{C}S_r(E), \nonumber \\
\frac{\partial \bF_r}{\partial t}+\mathbb{C}\bfnabla\cdot{\sf P}_r&=&\mathbb{C}{\bf S_r}(\bP),
\label{equations}
\end{eqnarray}
where the dimensionless source terms are,
\begin{eqnarray}
{\bf S_r}(\bP)&=&-\sigma_t\left(\bF_r-\frac{\bv E_r+\bv\cdot{\sf P} _r}{\mathbb{C}}\right)+\sigma_a\frac{\bv}{\mathbb{C}}(T^4-E_r) ,\nonumber\\
S_r(E)&=&\sigma_a(T^4-E_r)+(\sigma_a-\sigma_s)\frac{\bv}{\mathbb{C}}\cdot\left(\bF _r-\frac{\bv E_r+\bv\cdot{\sf  P} _r}{\mathbb{C}}\right),
\label{sources}
\end{eqnarray}
while $\sigma_a$ and $\sigma_s$ are the absorption and
scattering opacities. Total opacity (attenuation coefficient) 
is $\sigma_t=\sigma_s+\sigma_a$.  
The dimensionless number $\Prat\equiv a_rT_0^4/P_0$ is a measure of 
the ratio between radiation pressure and gas pressure in the fiducial units. 
We prefer the dimensionless equations because the dimensionless numbers, 
such as $\Crat$ and $\Prat$, can quantitatively indicate the importance 
of the radiation field as discussed in \cite{Jiangetal2012}.

We solve these equations in a 2D $x-y$ plane with the recently developed 
radiation transfer module in Athena \citep[][]{Jiangetal2012}. The continuity 
equation, gas momentum equation and gas energy equation are solved with 
a modified Godunov method, which couples the stiff radiation source terms to 
the calculations of the Riemann fluxes. The radiation subsystem for $E_r$ 
and $\bF_r$ are solved with a first order implicit Backward Euler method. 
Details on the numerical algorithm and tests of the code are described 
in \cite{Jiangetal2012}. The variable Eddington tensor is computed from angular 
quadratures of the specific intensity $I_r$, which is calculated from 
the time-independent transfer equation 
\begin{eqnarray}
\frac{\partial I_r}{\partial s}=\kappa_t (S-I_r).
\label{calEdd}
\end{eqnarray}
Details on how we calculate the VET, including tests, are given in \cite{Davisetal2012}.

Most of our simulations are performed using 
our standard computational domain 
$(x_{min}, x_{max}) \times (y_{min}, y_{max})$ which is
$(-0.5,0.5) \times (-0.5,0.5)$~ and standard resolution which is
$512\times512$ cells. 
Initially the background medium has
density $\rho_0=1$ $\rm g/cm^3$ and temperature $T_0=10^6$~K.  An over-dense
clump is located in a circular region $r\equiv x^2/x_0^2+y^2/y_0^2\le
1$, with $x_0=y_0=0.05$~cm.  The density inside this
region is $\rho(x,y)=\rho_0+(\rho_1-\rho_0)/[1.0+\exp(10(r-1))]$, where
$\rho_1$ is a free parameter.  The clump is in pressure equilibrium
with its surroundings, so the interior is colder than the
ambient medium.  The initial radiation temperature is the same as
the gas temperature everywhere.  Here we consider opacities due to scattering, $\sigma_s=\sigma_{s,0}(\rho/\rho_0)$, 
or absorption $\sigma_a=\sigma_{a,0}(T/T_0)^{-3.5}(\rho/\rho_0)^2$~$\rm cm^{-1}$, 
or both $\sigma_t=\sigma_s+\sigma_a$.
The radiation flux $\bF_r$ is zero everywhere initially.  We use reflection 
boundary conditions on both $y$ boundaries, outflow boundary conditions on 
the right $x$ boundary. A constant radiation field with 
temperature $T_r=2T_0$ is input through the left $x$ boundary.
At the left $x$ boundary, the gas temperature and
density are fixed to $T_0$ and $\rho_0$ respectively.  
The dimensionless speed of
light $\Crat=3.3\times10^3$, and the parameter $\Prat=10^{-3}$.

\section{Results}
\label{sec:results}

We have performed over thirty different simulations exploring the
parameter space and numerical effects. Here we discuss in some detail 
five simulations that illustrate the evolution of a  cloud in significantly 
different physical regimes: a optically thin and thick cloud with 
the pure scattering opacity (runs S10 and S200, respectively) and 
mildly optically thick, optically thick, and very optically thick cloud with
absorption dominated opacity (runs A10, A40, and A80, respectively).
Our convection of naming the simulations is the following: the letter A 
or S stands for the opacity type, i.e., dominated by absorption or scattering, 
respectively. A number that follows the letter corresponds to 
the initial cloud density, $\rho_1$.

We summarize the properties of the five simulations in
Table~\ref{table:summary}.
Columns (2), (3) and (4) give the input physical parameters:
$\sigma_{a,0}$, $\sigma_{s,0}$, and $\rho_1$. 
In columns (5), (6), (7), (8), and (9)
we list the following initial properties of the cloud:
the absorption optical depth, $\tau_{a}=2 x_o \sigma_a$,
the scattering optical depth, $\tau_{s}=2 x_o \sigma_s$,
the radiation diffusion time across the cloud,
$t_{dif}=4x_o^2\sigma_t/\Crat=\tau_t t_{fs}$, where $t_{fs}=2x_o/\Crat$ is 
the free-streaming time\footnote 
{$t_{fs}=6.06\times10^{-5}$ and it is the same for all the cases discussed.},
the thermal time scale inside the cloud, $t_{th}=P/(\Prat \Crat E_r \sigma_a)$,
and the sound crossing time
$t_{sc}=2 x_o/c_s$, where $c_s=\sqrt{\gamma P/\rho}$
is the adiabatic sound speed. 
Finally, columns (10), (11), and (12) give 
the numerical resolution, $n_x \times n_y$,
the Courant number, $C_0$, 
and the final time at which we stopped each simulation, $t_f$.
For the five runs, we used our standard computational domain.

\begin{figure*}
\begin{sideways}
\put(-140,370){\includegraphics{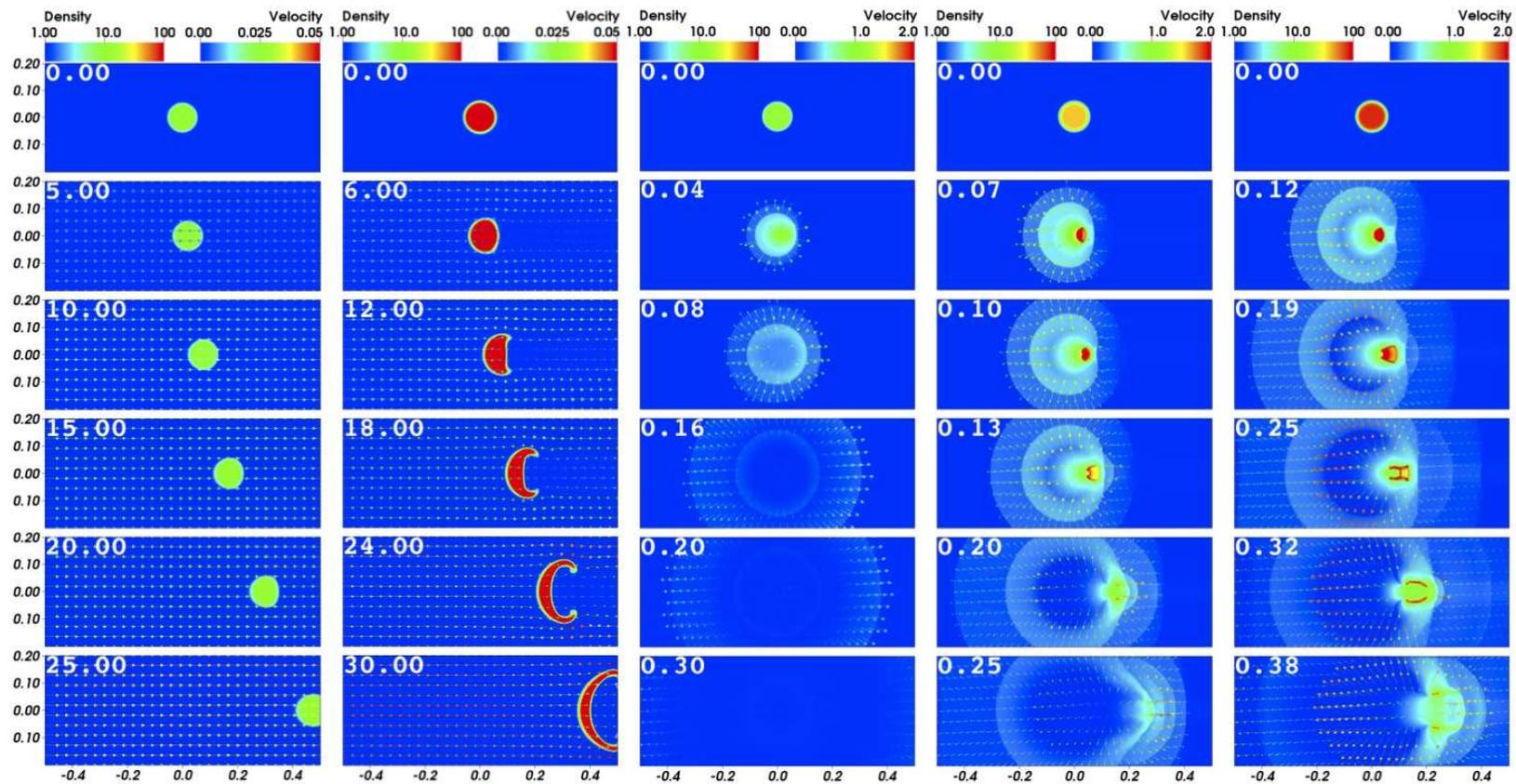}}
\caption{Sequences of density maps overplotted by the velocity fields
  for runs S10, S200, A10, A40 and A80 (from left to right) at five different
  times shown at the top left corner of each panel (from top to bottom)}
\label{fig:gallery}
\end{sideways}
\end{figure*}

Figure~\ref{fig:gallery} gives an overview
of the cloud evolution in the five runs (columns from left
to right correspond to runs: S10, S200, A10, A40, and A80) .
Specifically, the figure shows sequences of density maps overlaid
with velocities at five different times (the
time increases from top to bottom; the actual
time is given in the top left corner in each panel).

\begin{figure*}[ht]
\centering
\begin{sideways}
\put(-80,400){\includegraphics{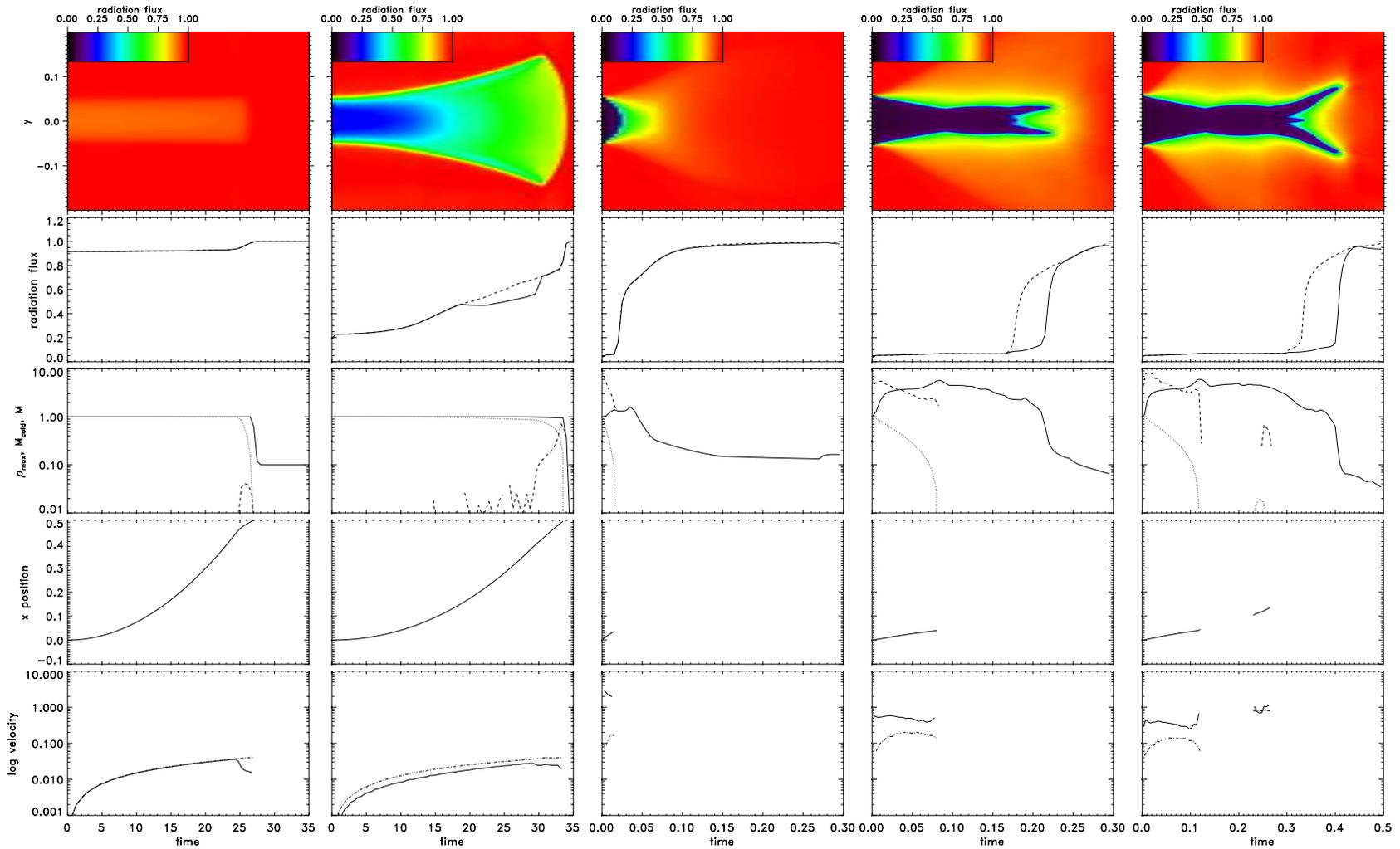}}
\caption{Time evolution of various properties for each run 
(from left to right: run S10, S200, A10, A40, and A80).
{\it From top to bottom:} 
the $x$ component of the radiation field, $F_{r1}$ as a function of $y$
along the right boundary of the computational domain (top panels);
the minimum value of $F_{r1}$ and $F_{r1}$ at $y=0$ along the right edge
of the computational domain, the solid and dashed lines, respectively
(second top panels);
the total mass and mass loss rate of the cold gas and the maximum density, 
the dotted, dashed, and solid lines, respectively (middle panels);
the $x$-position of the center of mass of the cold gas (second bottom panels);
the $x$ velocity of the center of mass and the maximum velocity of the
cold gas, the solid and dash-dotted lines (bottom panels).
The fluxes in the top and second top panels are normalized so that they 
are in units of the maximum radiation flux along the right boundary
at a given time.
}
\label{fig:time}
\end{sideways}
\end{figure*}

Figure~\ref{fig:time} shows several cloud
properties as a function of time. The figure
illustrates more qualitatively the differences
between various runs 
(the columns' correspondence to various runs is the same as 
in Fig.~\ref{fig:gallery})
as well as dramatic changes in the clouds with time.

To illustrate how a given cloud would appear to an observer
measuring the radiation at the right side of the computational
domain, the panels in the top row present
the $x$ component of the normalized radiation field, $F_{r1}$, 
as a function of the $y$ coordinate. 
The second from  top row of panels show $F_{r1}$ at the right boundary
but only at $y=0$ (i.e, the $(x_{max},0)$ location) and its minimum value 
along the right boundary (the dashed and solid lines, respectively). 
The fluxes in these panels are normalized so that they are in units of 
the maximum radiation flux along the right boundary at a given time.
The middle panels present the time evolution of the {\it cold} gas
in the computational domain:
the maximum density, total mass and
the mass loss rate of the cold gas 
(the solid, dotted, and dashed lines, respectively).
We normalized the maximum density to the initial density of the cloud, 
$\rho_1$. Our operational definition of ``cold'' gas
is the gas with the temperature less than 2 times 
the initial temperature of the cloud.
Therefore, the middle row panels can be used to follow the cloud heating and
subsequent evaporation.

To follow the average cloud motion, the second bottom row of panels show 
the $x$-position of the center of mass (CoM) of the cold gas, while
the bottom panels show
the $x$ component of the CoM velocity and the maximum velocity of the
cold gas (the solid and dash-dotted lines, respectively).

We start by discussing run~S10 that is related to one 
of the simple cases mentioned in \S~\ref{sec:scases}: simple example~II,
with scattering opacity only 
and relatively small optical depth (i.e., $\tau_s=0.2$).
The left columns of panels of Figs.~\ref{fig:gallery} 
and ~\ref{fig:time} show that,
as expected, there is no compression (i.e., $\rho_{max}=1$)
no loss of the cold gas, and no shock formation. The 
cloud is almost uniformly accelerated in the horizontal direction.
The implied mass loss of cold gas for $t \geq 27$ is simply the result of the 
cold cloud being advected out of the domain.
In addition,  the cloud is almost co-moving with the ambient medium
We note that the acceleration due to the radiation force, $a_{rad}$
is almost position and time independent in this optically
thin case, i.e., 
$a_{rad}=\mathbb{P} \sigma_{s,0} F_{r1}\simeq 1.6\times10^{-3}$
(one can ignore the velocity dependent terms in the source term,
${\bf S_r}(\bP)$, in the momentum equation as these terms are very small, i.e.,
$v/\mathbb{C}<<1$).
This acceleration is relatively small. Specifically, the time for an optically
thin fluid element to travel a distance equal to the cloud diameter,
$t_{dyn}=(4 x_o/a_{rad})^{1/2}\approx 11$ is long
compared to $t_{sc}=0.24$.

Also as expected, the position of CoM is a quadratic function of time 
and the maximum velocity of the cold gas is very similar to the velocity of
CoM (see the two lowest panels of the left hand side column in 
Fig.~\ref{fig:time}).
The small but not zero optical depth means that the front side
of the cloud experiences a slightly stronger  push by radiation
than the back side (the acceleration at the back compared
to that at the front is smaller by a factor of $0.9$).
Consequently, the cloud is flattened by radiation.
This effect is much stronger in run~S200, where the cloud
is optically thick.

The second left column of panels in Figure~\ref{fig:gallery}
illustrate how the radiation initially 'squeezes' the cloud in S200: 
the front side is pushed by radiation while the back side does not move.
The radiation pressure acts only from the left side and the gas pressure
inside the cloud is higher than the pressure of the ambient gas.
This pressure imbalance leads to lateral expansion that is noticeable 
already by the time, $t=12$. The cloud evolution is further complicated
by the fact that the optical depth decreases (not always monotonically)
as a function of $y$
from the cloud center to its edge. For example,
as the cloud moves as a whole to the right
it also expands laterally and its shape 
starts to resemble a crescent because the more transparent
edges are pushed more than the cloud center. 

The top panel in the second left column in Figure~\ref{fig:time}
shows clearly that the shadow size increases with time. However, 
the panel also shows that at the later times (i.e., $t\simgreat20$)
the cloud center is not the most opaque
part of the cloud. Instead the most opaque regions are near the edges where
the  horns of the crescent bend over toward the center
(the second top panel  shows this too: for $t\simgreat20$, 
the minimum radiation field
along the $y$ direction at $x=x_{max}$ is not at $y=0$).

Overall the cloud in run~S200 moves slower than in run~S10
(compare the second to bottom and bottom panels in corresponding columns
in Fig.~\ref{fig:time}). In addition, for run~S200, 
the CoM velocity is lower than the maximum velocity of the cold gas. 
This difference in the velocities is an indication
of a non-uniform cloud evolution. The two just presented examples
show that even for a pure scattering opacity case, the cloud
evolution is fast and very different than
the movement of a bullet where there is no change in the shape and size
as the cloud moves. The evolution of a cloud with absorption dominated opacity
is even more dynamic and complex and also faster.

We start by discussing run~A10 that is related to another 
simple case mentioned in \S~\ref{sec:scases}: simple example~I - the cloud
behaving as a balloon. 
In this run, $\tau_a=2\times10^3$, $t_{dif}=1.2\times10^{-1}$,
while $t_{th}=2\times10^{-6}$. Thus, both $t_{dif}$ and $t_{th}$ are much 
smaller than $t_{dyn}$. Therefore, the evolution of the absorption dominated
cases is much faster than the pure scattering cases. For example,
the middle column of panels of Fig.~\ref{fig:gallery}
show that the cloud expands within time less than 0.2.
During its evolution, the cloud does not gain much net momentum and 
it remains almost spherical. 

The total mass of the cold gas  drops to zero within $t=3\times10^{-2}$ 
(see the middle panel in the middle column in Fig.~\ref{fig:time}).
This time is of the order of $t_{dif}$.
The gas density in the cloud is initially increased by a factor of 2 but
for $t\simgreat0.05$ $\rho_{max}$ 
decreases below 0.01. We note that in this run,
the velocity of CoM does not correspond to the movement
of the whole cloud but rather to the change in the location
of the boundary between the cold and hot
(to the rapid propagation of TF).
Hence, the implied CoM velocity can exceed the maximal velocity in 
these runs. 
The bottom panel in the middle column of Figure~\ref{fig:time}
shows that for as long as the cold gas exists,
its velocity increases very fast with time.
After $t_{dif}$, the only gas left on the grid is 
the transparent and hot gas moving almost radially away
from the center of the grid
where the initial cloud was located.

The cloud in run~A10 is initially optically thick. However,
once the fast TF passes the cloud, the optical depth drops to 
about 0.6 due to an increase of the temperature alone (i.e.,
with small change in the cloud density). This quick, isochoric decrease
in the cloud optical depth is the main reason for 
the evolution in run~A10 to resemble the evolution
of simple example~I which we referred to as an expanding balloon.
Runs with $\tau_a$ lower than in run~A10 show very similar evolution 
but occurring on a longer time scale. However runs with higher optical depth
show substantial qualitative differences. A good indicator
of the expected difference in the evolution is the relation between
the diffusion and sound crossing times.

For run~A10, $t_{dif}< t_{sc}$ and the cloud behaves like a balloon
even though it is initially optically thick.
However, for run~A40 with $t_{dif}> t_{sc}$ (and $t_{th}$ shortened
by almost 4 orders of magnitude), we observe strong evaporation
from the irradiated side, the development and propagation of
a strong shock inside the cloud, and a few  other features unseen in run~A10
(compare the middle and second right columns of panels
in Figs.~\ref{fig:gallery}~and~\ref{fig:time}).

In run~A40, the radiation heats the front much faster than it can diffuse 
across 
the cloud. Therefore, the cloud loses mass from the left side through 
a hot outflow. In addition, a very fast and strong shock propagates to 
the right of the cloud compressing it and changing its shape.
The changes in the shape are major: from the initial convex shape through 
the concave one of the front side to the break up of the cloud 
into two smaller elongated clumps.

The cloud evolution is far from being isochoric. For example,
the density on the left side of the cold part of the cloud is increased 
by a factor of 8 (i.e., $\rho_{max}=8)$ at $t=0.9$. 
This time corresponds to the maximum compression 
of the cloud and the smallest shadow (see the fourth column of panels in  
Figs.~\ref{fig:gallery} and ~\ref{fig:time}). After this time the cloud 
re-expands and fragments (around $t=0.17$). Finally, at $t\approx 0.25$, 
even the dense clumps become optically thin as they are heated and dispersed.

As expected, the cloud in run~A40 survives much longer than in run~A10.
Another consequence of the higher cloud density is that
the dense parts of the cloud can travel over some distance before they are 
heated and dispersed. For example, the second bottom panel in the fourth
column of Fig.~\ref{fig:gallery} shows two dense and optically thick clumps
at $x=0.18$ for t=0.20 -- more than three radii of the initial cloud away
from the original location. This cloud movement to the right is caused by the 
cloud hot outflow that moves to the left (i.e., the rocket effect).
The push by radiation pressure in this run as well as in runs A10 and A80 is
negligible because $t_{dyn}\approx 11$ which is orders of magnitude
longer than $t_{th}$.

To show the evolution of an even denser cloud, we carried out a simulation 
for $\rho_1=80$ (run~A80). The right hand side panels of 
Figs.~\ref{fig:gallery} and ~\ref{fig:time} show the results of this 
simulation. Overall, the evolution of this denser cloud resembles that of 
the cloud in run~A40. However, there are some quantitative differences.
In particular, in run~A80, the cloud stays optically thick for about 
twice as long as in run~A40  (compare the second top panels in the second right
and right columns in Fig.~\ref{fig:time}). The two clumps that form after 
the re-expansion phase are denser and more elongated. 

The middle to bottom panels in the right column of the figure suggest that 
the cloud disappears at $t\approx0.13$ and reappears around  0.24. However, 
this is an artifact of our formal definition of the cold phase (the gas with 
the temperature less than 2 times the initial temperature of the cloud).
The 'disappearance' of the cold phase corresponds the smallest cloud size 
with the gas temperature increased due to compression rather than radiative 
heating. However, as the dense cloud later re-expands, the gas temperature 
drops below our temperature threshold. Later still, 
the cloud moves to the right, fragments and continues to be heated
by radiation so that  at $t\approx 0.40$ it is totally dispersed.

\section{Summary and Discussion}
\label{sec:summary}

We have presented two-dimensional, radiation-hydrodynamic calculations
of time-dependent structure of  strongly irradiated clouds. 
Our primary conclusions are the following:
1) even a relatively dense cloud that is radiatively heated 
(i.e., with significant absorption opacity) does not move as a whole, and 
instead undergoes a very dramatic evolution in its shape, size and physical 
properties. In particular, the cloud and its remnants become optically thin 
within less than one sound crossing time and 
before they can travel over a significant distance  (a distance of a few 
radii of the initial cloud) and
2) a cloud can be accelerated as a whole under quite extreme conditions, e.g.,
the opacity must be dominated by scattering. However, the radiation force
acceleration is relatively small and unless the cloud
is optically thin, the cloud quickly changes its size and shape.

A competition of several physical processes determines the cloud evolution.
In the cases that we have explored, the most important processes 
are the radiation diffusion and heating, the radiation force acceleration,
and finally propagation of the sound speed. 
Comparing the time scales corresponding
to each of these processes (see \S~\ref{sec:intro}),
one can predict the behaviour of a cloud with given initial
conditions. For example, as we noted in \S~\ref{sec:results},
for absoprtion dominated opacity and
$t_{dif} < t_{sc}$, a cloud will behave like a balloon (simple example~I)
even if the cloud is initially optically thick.
The radiative heating is usually the fastest process. Therefore,
simple example~II (a bullet like case) requires very special conditions:
scattering dominated opacity and the optical depth much less than one.

Our goal here was to carry out simulations in a simple and well-controlled way 
so that we can assess the role of the type and magnitude of the opacity on the 
cloud evolution. While achieving this goal, we found that Athena can handle well 
quite a wide range of physical conditions. Not surprisingly, we also found that 
the most difficult cases to compute are those with huge density and opacity 
contrasts,  especially when the opacity is dominated by absorption. In these cases,
the thermal time scale can be extremely short so that the equation of energy is 
stiff. To avoid spurious oscillations, we had to increase the numerical resolution 
and reduce the Courant number (see columns 10 and 11 in Table~\ref{table:summary}).
For example, in run~A80, we reduced $C_0$ to 0.02 from its usual value of 0.8.

In the runs discussed here, we increased the temperature of the incoming radiation 
by a relatively small factor of 2 ($T_r=2 T_0$). This corresponds to an increase 
of the radiation pressure by a factor of 16. Results from our  various tests with 
higher $T_r$ and also higher $\Prat$ confirm simple expectations that increasing 
$T_r$ or $\Prat$ results in a faster cloud evolution because of the increases
in the heating rate and cloud acceleration. 

As we mentioned in \S~\ref{sec:intro}, the clouds in the BLRs are optically thick 
to absorption and optically thin to scattering. Therefore, the results from 
runs A40 and A80 are most applicable to the BLRs of AGN. These results well 
illustrate the fact that any cloudy model for the BLRs faces the issues 
of stability and confinement. 

These are very serious issues because in our simplified simulations, the cloud
velocity was not very different from the ambient velocity. 
In particular, we did not assume that the cloud is on a Keplerian
orbit around the central black hole as it is assumed in some BLR models 
\citep[e.g.,][and references therein]{Pancoastetal2011,Krauseetal2012}.
Therefore, here the cloud was not much affected by hydrodynamic instabilities 
caused by velocity shear. Nevertheless, we found that the cloud can not
be significantly accelerated before it is dispersed (i.e., the rocket effect 
is not efficient). The main process responsible for the dispersal is radiative 
heating. One of the implications of these results is that clouds in the BLRs, 
if they exist, likely take on many different and complex shapes and this 
complication should be considered in detailed calculations of cloud line emission.

There are a number of limitations to our results. Probably the most important are 
that we neglected gas photoionization and assumed frequency-independent (gray) 
opacities. Additional limitations to our calculations are that they are 2-D 
instead of fully 3-D and that we did not include thermal conduction, 
magnetic fields, and the process or processes responsible for the cloud formation. 
Moreover, in realistic simulations of the BLRs, one should also consider computing 
the evolution of interacting multiple clouds with various properties, not just 
one cloud as we did here. Our results indicate that such multi-cloud 
simulations are feasible using Athena.

We plan to go beyond some of these limitations in the near future. 
In particular, we plan to address the issue of cloud formation and subsequent
evolution. One possible approach would be to consider the radiative processes
and conditions for which the gas is thermally unstable so 
that cold condensations can grow on relatively short time scales
despite the presence of a strong radiation field
\citep[e.g.,][and references therein]{Krolik1988, MoscibrodzkaProga2013}.
Such an approach, our intended next step,
will involve simulations quite different from those presented 
here because among other things, here the gas is 
thermally stable and initially it is not in a radiative equilibrium.

\section*{Acknowledgments}

This work was supported by NASA under ATP grant NNX11AI96G and NNX11AF49G. 
D.P. thanks S. Lepp for discussions and also the Department of Astrophysical 
Sciences, Princeton University for its hospitality during his sabbatical
when the work presented here was initiated. He also
wishes to acknowledge the National Supercomputing Center
        for Energy and the Environment, for providing computer resources
        and support.
Y.F.J. was provided by NASA through Einstein Postdoctoral Fellowship grant
number PF-140109 awarded by the Chandra X-ray Center, which is operated by
the Smithsonian Astrophysical Observatory for NASA under contract NAS8-03060.
\bibliographystyle{apj}
\bibliography{cl}

\clearpage
 
\begin{table*}[ht]
\scriptsize
\begin{center}
\caption{ Summary of runs}
\begin{tabular}{l c c  c c  c c c c c c c  } \\ \hline
 & & & & & & & & & & \\
Run &$\sigma_{a,0}$ & $\sigma_{s,0}$ & $\rho_{1}$ & $\tau_{a}$ & $\tau_{s}$ &
$t_{dif}$ & $t_{th}$ & $t_{sc}$ &$n_{x}\times n_{y}$  & $C_o$   & $t_f$ \\ \hline
 & & & & & & & & & & \\
S10     &  0  & 0.1 & 10  & 0 &   0.2 & - &
$\infty$ & 0.48 & $512\times512$ & 0.8    &
50 \\ 
S200   & 0  & 0.1 & 200  & 0 & 4.0 & $2.4\times10^{-4}$ &
$\infty$ & 2.20  & $512\times512$ & 0.8 & 50     \\  
 & & & & & & & & & & \\
A10   & $3.16\times10^{-2}$  & 0.1 & 10  & $2\times10^3$ & 0.2
&$1.2\times10^{-1}$ & $2\times10^{-6}$ & 0.48 & $1024\times1024$ & 0.1 & 0.3 \\  
A40   & $3.16\times10^{-2}$  & 0.1 & 40  & $4\times10^6$
&0.8 & $2.4\times10^2$ & $9\times10^{-10}$ & 0.98 &$1024\times1024$ & 0.1   & 0.3 \\  
A80   & $3.16\times10^{-2}$  & 0.1 & 80  & $1.8\times10^8$ &
1.6 & $1.2\times10^4$ & $2\times10^{-11}$ & 1.38 & $1024\times1024$ & 0.02 & 0.6 \\  \hline

\end{tabular}~\label{table:summary}
\end{center}
\normalsize
\end{table*}


\end{document}